Estimation of the surface mechanical properties of soft tissues mimicking phantoms using impact analyses: a comparative study

# Estimation of the surface mechanical properties of soft tissues mimicking phantoms using impact analyses: a comparative study

**Arthur Bouffandeau[1], Sabine Bensamoun[2], Robert Schleip[3,4,5], Giuseppe Rosi[6], Charles-Henri Flouzat-Lachaniette[7,8], Jean-Paul Meningaud[8,9] and Guillaume Haiat[1]**

**[1]CNRS, Univ Paris Est Créteil, Univ Gustave Eiffel, UMR 8208, MSME, Créteil, France.**

**[2]Université de technologie de Compiègne, CNRS, BMBI (Biomechanics and Bioengineering), Compiègne, France.**

**[3]Conservative and Rehabilitative Orthopedics, Department of Sport and Health Sciences, Technical University of Munich, Munich, Germany.**

**[4]Department for Medical Professions, Diploma Hochschule, Bad Sooden-Allendorf, Germany.**

**[5]Experimental Anaesthesiology, Ulm University, Ulm, Germany.**

**[6]Univ Paris Est Créteil, Univ Gustave Eiffel, CNRS, UMR 8208, MSME, Créteil, France.**

**[7]Service de Chirurgie Orthopédique et Traumatologique, Hôpital Henri Mondor AP-HP, CHU Paris 12, Université Paris-Est, Créteil, France.**

**[8]INSERM U955, IMRB Université Paris-Est, Créteil, France.**

**[9]Aesthetic and Maxillofacial Surgery Department, Hôpital Henri Mondor, Plastic, Reconstructive, Créteil, France.**

Estimation of the surface mechanical properties of soft tissues mimicking phantoms using impact analyses: a comparative study

## Abstract

**Background –** Palpation is the most widely used approach to empirically assess the mechanical properties of superficial tissues. While elastography is used for volume measurements, it remains difficult to assess skin properties with non-invasive methods. This study aimed to compare the performances of an impact-based analysis method (IBAM) consisting in studying the dynamic response of a punch in contact with the tissue with other approaches available on the market.

**Materials and Methods –** IBAM consists in analyzing the time dependent force signal induced when a hammer instrumented with a force sensor impacts a cylindrical punch placed in contact with soft tissue. Sensitivities to stiffness changes and to spatial variations were compared between IBAM and four other mechanical surface characterization techniques: IndentoPro® (macroindentation), Cutometer® (suction), MyotonPro® (damped oscillation) and Shore Durometer (durometry) using soft tissue phantoms based on polyurethane gel.

**Results –** For stiffness discrimination in homogeneous phantoms, IBAM was slightly better than IndentoPro and MyotonPro (by 20 % and 35 % respectively), and outperformed the Shore Durometer and Cutometer by a factor of 2 to 4. Furthermore, for stiffness and thickness variations in bilayer phantoms, the axial sensitivity of IBAM was between 2.5 and 4.5 times better than that of MyotonPro and IndentoPro. In addition, the Cutometer appeared to be severely limited by its measurement depth.

**Conclusion -** IBAM seems to be a promising technique for characterizing the mechanical properties of soft tissue phantoms at relatively low depth. These results will need to be confirmed in future *in vivo* measurements on biological tissues. This work could pave the way to the development of a decision support system in the field of dermatology and cosmetics. However, its clinical applicability remains to be demonstrated *ex vivo* and *in vivo*.

# 1. Introduction

The mechanical properties of soft tissues depend on their pathophysiological conditions [1] and can be altered by various pathologies such as cancer, wound healing or musculoskeletal disorders [2–4]. Palpation, is commonly used clinically to empirically characterize soft tissue conditions [5] but it remains highly dependent on the experience of the clinician, has relatively low sensitivity to soft tissue stiffness, and cannot be used for follow-up. Consequently, various tools capable of characterizing the biomechanical properties of soft tissues have been developed. A user-friendly, non-invasive, and objective technique could become a valuable decision-support system in multiple medical domains, such as dermatology and plastic surgery [6,7]. Such approach could pave the way for personalized medicine, enhancing diagnosis, monitoring, and treatment evaluation. Additionally, skin characterization is also of great interest for the cosmetics industry [8,9], because it could allow to measure product efficacy and offer customer-specific products.

Using various medical imaging modalities, several elastographic techniques (static [10], continuous dynamic [11] and transient dynamic [12]) have been developed to estimate soft tissue stiffness for more than 30 years [5,13] with some techniques being used in the clinics [12,13]. However, the relative complexity of this technique and its cost still prevent its use in all clinical facilities [14]. Moreover, it still remains difficult to use current elastographic techniques to evaluate the skin properties *in vivo* [15,16]. Consequently, new relatively low-cost, simple and non-invasive devices have been developed to characterize the superficial biomechanical properties of soft tissues. The Cutometer® is a reference method in cosmetics that exploits the principle of suction to estimate the skin viscoelastic properties [6,9]. The Durometer is used in mechanical engineering to measure the hardness of several materials such as rubber, elastomers and plastics [17,18]. The MyotonPro® is a hand-held research device that measures the dynamic biomechanical properties and has already demonstrated its performances for the characterization of various soft tissues such as skeletal muscles [14,19], tendons [14] and skin [7,20]. The IndentoPro® is an indentometry method inspired by the Semi-electronic Tissue Compliance Meter (STCM) [21] and developed for the study of myofascial tissues [22,23]. Reliability, depth of characterization and sensitivity are different for each technique and should be taken into consideration.

More recently, a new Impact-Based Analysis Method (IBAM) was developed by our group for the mechanical characterization of the surface of soft tissues mimicking phantoms [24,25]. This method consists in analyzing the force signal as a function of time induced by the impact of a hammer instrumented with a force sensor on a cylindrical punch placed in contact with soft tissue. The IBAM method derives from previous works of our group performed in the context of orthopedic surgeries. An impact analysis technique was developed to quantitatively evaluate the stability of cementless hip arthroplasty implants: the acetabular cup [26–28] and the femoral stem [29–31]. This technique was then extended to osteotomy procedures in order to provide feedback to the surgeon on the mechanical information of the bone tissue surrounding the osteotome tip [32–36].

The sensitivity of IBAM to variation in the agar concentration of soft tissue phantoms was estimated and revealed as being similar to that of elastographic techniques [24]. Then, a previous experimental version of IBAM was compared with a digital palpation device, MyotonPro, using agar-based soft tissue phantoms [25]. IBAM demonstrated better performances than MyotonPro® in terms of sensitivity to stiffness and to axial variations and similar performances in terms of sensitivity to stiffness compared to dynamic mechanical analysis (DMA) [25]. However, in these previous studies, the variability and the fragility of the agar-based phantoms as well as the manual gesture were major limitations of IBAM.

The aim of this paper is to evaluate the capabilities of an improved version of IBAM to determine the biomechanical properties of soft tissues mimicking phantoms. The IBAM method was improved by standardizing the impact motion. Here, we focus in particular on the sensitivity of IBAM to determine i) changes of stiffness of homogeneous phantoms and ii) axial variation of heterogeneous phantoms. The performances of IBAM are compared with those of four other mechanical surface characterization techniques: IndentoPro®, Cutometer®, MyotonPro® and Shore Durometer. To replicate the mechanical behavior of soft tissue, polyurethane gel-based phantoms of different thicknesses and stiffnesses were considered, allowing to obtain more reproducible and standardized samples.

## 2. Materials and methods

### 2.1. Soft tissues mimicking phantoms

In this comparative study, polyurethane gel plates covered with a 25 µm thick polyurethane film (Technogel Germany GmbH, Berlingrode, Germany) were used to mimic the mechanical properties of soft biological tissues [37]. Therefore, this study does not require any ethical approval because no animal or human tissue samples were used. These phantoms were used to reproduce the tissues of the human lumbar region [22].

All plate phantoms had the same dimensions of 300×210 mm. Their thickness and their shore values varied from 1 to 10 mm and from 20 to 65. The stiffness values of the gel phantoms were given by the manufacturer with the standardized scale shore OOO and we used Table 1 to convert them in Young's modulus [38].

| **Shore OOO** | 20 | 25 | 30 | 40 | 45 | 50 | 55 | 60 | 65 |
|---|---|---|---|---|---|---|---|---|---|
| **Young's modulus (kPa)** | 30.3 | 37.4 | 45.8 | 68.8 | 84.6 | 105 | 131 | 167 | 216 |

***Table 1:*** *Conversion of shore OOO values into Young's modulus* ***[38].***

## 2.2. Impact-based analysis method (IBAM)

The IBAM has been developed to characterize the biomechanical properties of superficial soft tissues [24,25]. As shown in Fig. 1, the two main components of IBAM are i) a 5 g impact hammer equipped with a force sensor (type 8204, Brüel and Kjær (B&K), Naerum, Denmark) and ii) a 4 mm diameter aluminum cylinder (referred to as "punch" in what follows) in contact with the tissue to be measured. During an acquisition, the hammer impacted the vertically guided punch positioned in contact with the soft tissue phantom to be characterized.

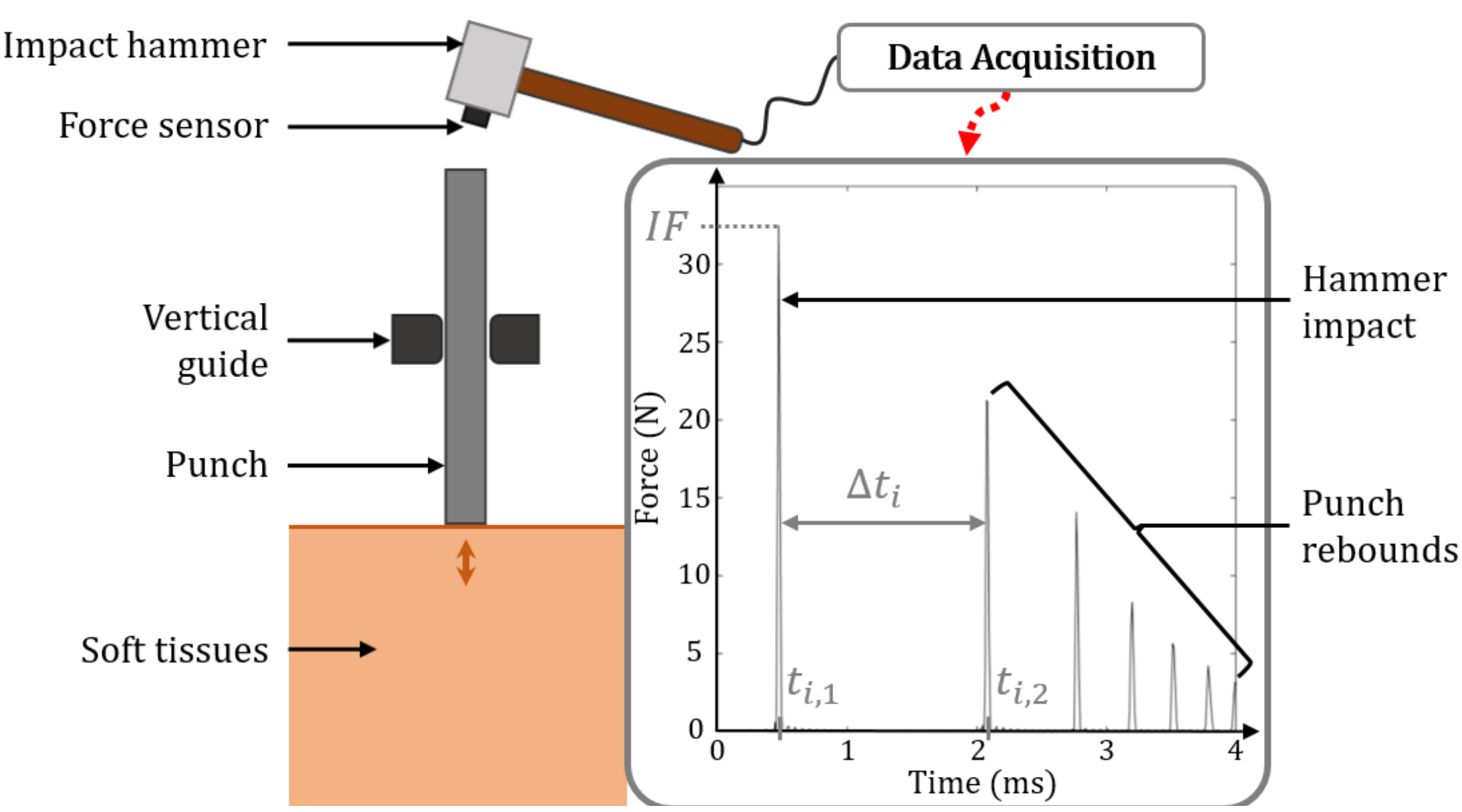

***Figure 1:*** *Left: illustration of the experimental set-up used by the impact-based analysis method (IBAM). Right: example of a temporal force signal with,* $\boldsymbol{\Delta t}_i$*=1.61 ms and* ***IF****=32.5 N.*

The experimental set-up described in Bouffandeau et al. [25] was improved by i) standardizing of the movement of the hammer using a fixed pivot which improves the reproducibility of the impacts, ii) optimizing the punch guidance, with new components such as a lifting platform, which simplifies the positioning of the punch relative to the sample, and reducing the friction to enhance punch guidance performance, and iii) in terms of the post-processing method with the implementation of an outlier detection method using the z-score metric with a threshold fixed at 2.5 [39].

The graph in Fig. 1 shows an example of the variation of the force as a function of time measured as output data of the force sensor using LabVIEW software (National Instruments, Austin, TX, USA) with a sampling frequency of 102.4 kHz (NI 9232, National Instruments, Austin, TX, USA). The time of the maximum value of the first peak (which corresponds to the initial impact of the hammer on the punch) was noted $t_{i,1}$. The amplitude of the first peak was noted $IF$ and corresponds to the impact force. The following peaks are related to the rebounds of the punch between the hammer and the phantom [24]. Thus, the time of the maximum value of the second peak (which corresponds to the first rebound) was denoted $t_{i,2}$. For each impact #$i$, the indicator $\Delta t_i = t_{i,2} - t_{i,1}$ was determined [24] and the indicator $\Delta t$, corresponding to the average of 5 consecutive impacts, was defined by:

$$\Delta t = \frac{1}{5}\sum_{i=1}^{5} \Delta t_i \quad (1)$$

Only the signals with an $IF$ value comprised between 30 and 35 N were considered in Eq. 1 for the determination of $\Delta t$. For each experiment, 5 measurements, each with 5 impacts, were realized to obtain the mean and standard deviation of the indicator $\Delta t$.

### 2.3. Mechanical surface characterization techniques

Four commercial devices using different physical principles to quantify soft tissue stiffness were chosen to reflect a wide spectrum of mechanical surface characterization techniques and provide a benchmark for comparison with the performances of the IBAM.

**IndentoPro** (Fascia Research Group, Ulm University, Department of Human Movement Sciences, University of Chemnitz, Germany) [22,23]. This hand-held digital device is based on an indentometry method to determine the stiffness of soft tissues. After selecting the indentation depth between 2 and 15 mm (2 mm for our measurements) and positioning the 11.3 mm diameter circular probe in contact with the tissue, the operator must apply a compressive force on the device, as shown in Fig. 2a. The output data is the stiffness value, noted $Ind$ given in N/mm, and correspond to the ratio between the applied force and the probe displacement, respectively recorded by a load cell and a membrane potentiometer. The value of $Ind$ increases as a function of the tissue stiffness. For each experiment, 5 measurements were realized to obtain the mean and standard deviation of the indicator $Ind$.

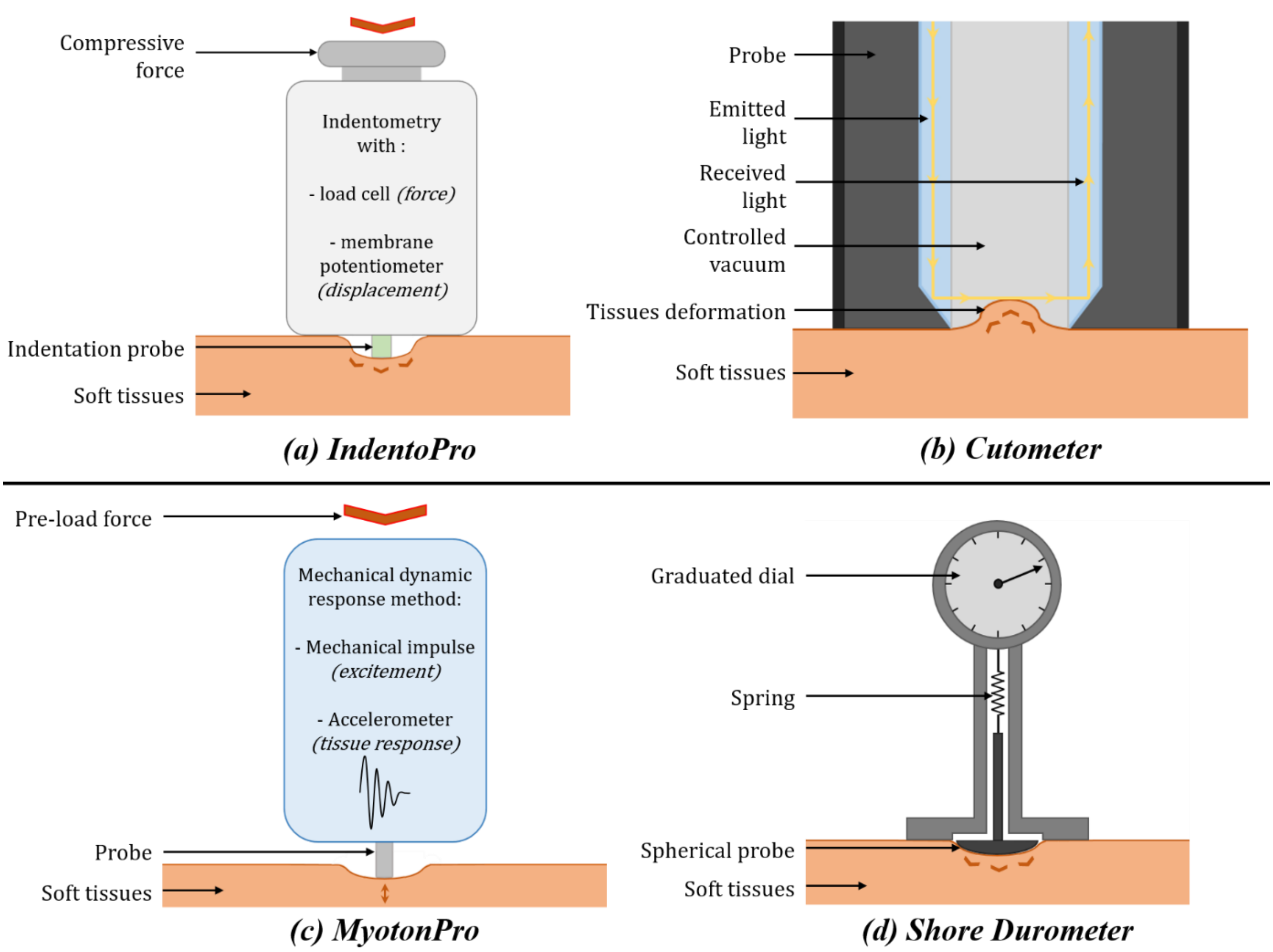

***Figure 2:*** *Illustration of the four commercial mechanical surface characterization techniques used as benchmark for the IBAM. (a) macroindentation using IndentoPro, (b) suction using Cutometer, (c) damped oscillation using MyotonPro and (d) durometry using Shore Durometer.*

**Cutometer** (MPA 580, Courage & Khazaka Electronic GmbH, Cologne, Germany). This device allows to assess the viscoelastic properties of the skin using the suction principle [6,8,9]. Cutometer is mainly used to evaluate the clinical condition of scars or the efficacy of cosmetic products. During a measurement, a controlled negative-pressure of 450 mbar is applied within the probe and enables to suck up the tissue through a 2 mm diameter aperture in the probe. A measurement is composed of 3 successive cycles of 2 s aspiration and 2 s relaxation. The variation of the tissue displacement as a function of time was recorded by optical sensors and then various parameters were extracted from this signal as shown in Fig. 2b. In our study, only the indicator $R_0$, which corresponds to the tissue deformation at the end of the aspiration cycle, was recorded and given in mm. The value of $R_0$ decreases as a function of the tissue stiffness. For each experiment, 5 measurements were realized to obtain the mean and standard deviation of the indicator $R_0$.

**MyotonPro** (MyotonAS, Tallinn, Estonia). This hand-held research device provides a dynamic estimation of the biomechanical properties of soft tissues such as tendons, muscles [14,19] and skin [7,20]. After positioning the device perpendicular to the tissue, a pre-compression force of 0.18 N was applied, followed by a mechanical impulse of 0.4 N for 7 ms [7], and repeated five times with a time interval of 0.8 s to perform a multi-scan measurement with the 3 mm diameter standard probe as shown in Fig. 2c [40]. The soft tissue response was recorded using an accelerometer and the system was modeled by a damped oscillatory system, allowing to extract several mechanical parameters. Only the dynamic stiffness $S$ (in N/m) was recorded herein. For each experiment, 5 measurements were realized to obtain the mean and standard deviation of the indicator $S$.

**Shore Durometer** (Type 1600-OOO, Rex Gauge, Lake Zurich, IL, USA), also called Durometer. This hand-held device is used to evaluate the hardness of non-metallic materials such as rubber, polymer [17] and soft tissue [7,18]. According to the international durometer standard (ASTM D2240), the hardness values are given according to shore scales, depending on the type of durometer used (in our case, type OOO), ranging from 0 to 100 degrees. The degrees of shore can be converted into Young's modulus [38], as shown in Table 1. The device must be positioned perpendicular to the soft tissue and the shore value $H$ was read from the graduated dial. This measured value was correlated to the indentation of the 12.7 mm spherical probe in the tissue due to the deformation of the spring (with a spring force of 113 g) as shown in Fig. 2d. Prior to the different series of experiments, a type OOO test block kit (Rex Gauge, Lake Zurich, IL, USA) was used to check the consistency and the proper handling of the device.

The value of $H$ increases as a function of the tissue stiffness. For each experiment, 5 measurements were realized to obtain the mean and standard deviation of the indicator $H$.

### 2.4. Comparative study protocols

The aim of this comparative study was to assess the performances in terms of sensitivity of the five mechanical surface characterization techniques (IBAM, IndentoPro, Cutometer, MyotonPro and Shore Durometer) using homogeneous and bilayer phantoms. Three distinct experimental protocols (Fig. 3) were considered and are described below.

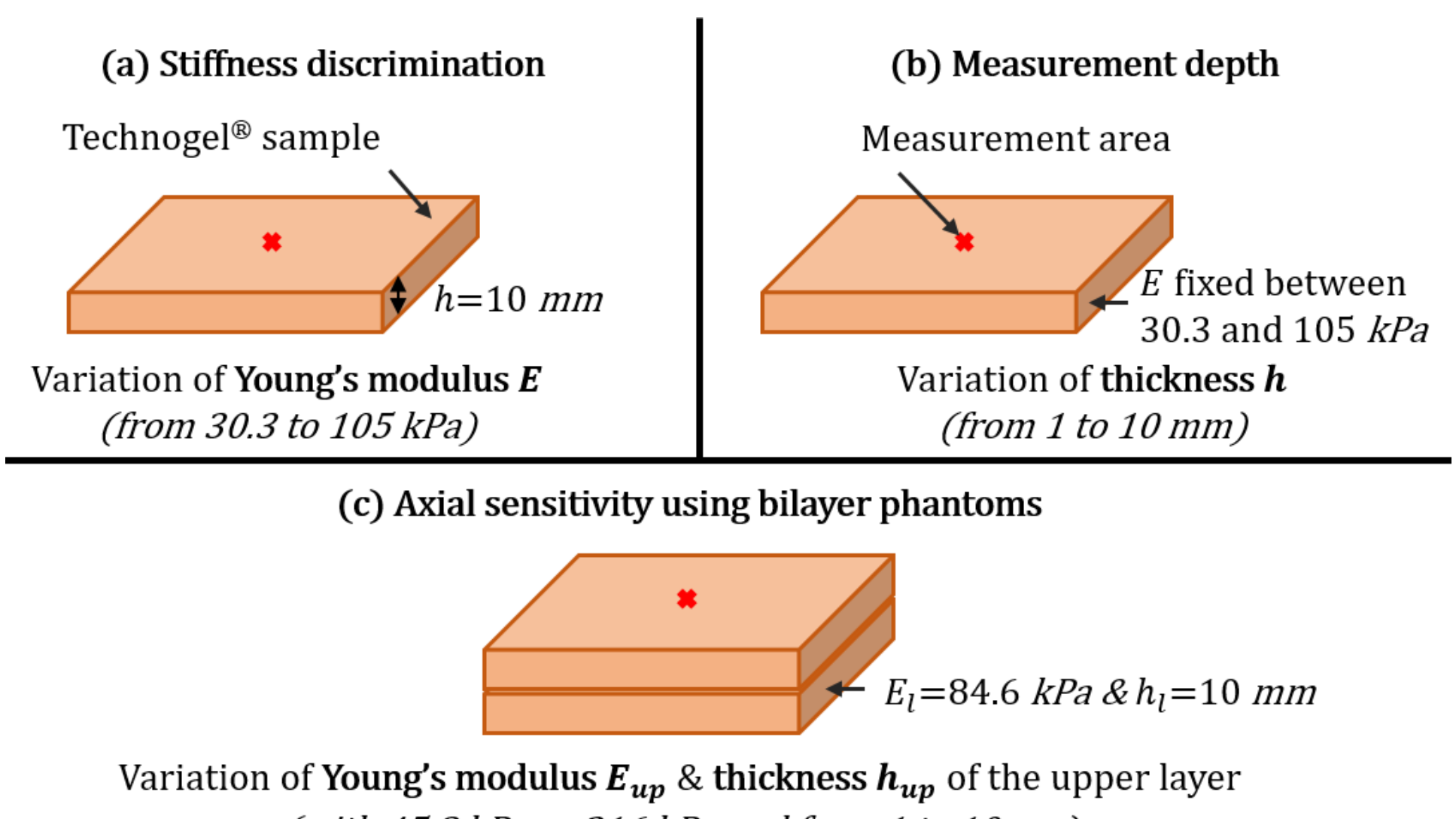

***Figure 3:*** *Illustration of the three protocols of the comparative study. (a) Evaluation of the stiffness sensitivity using homogeneous phantoms, (b) Estimation of the measurement depth using homogeneous phantoms and (c) Estimation of the axial sensitivity using bilayer phantoms.*

**Stiffness discrimination using homogeneous phantoms.** For this first protocol, the aim was to evaluate the effect of the Young's modulus on the values of the indicators for the five aforementioned techniques. For this purpose, six 10 mm thick polyurethane gel-based phantoms with Young's modulus $E$ varying from 30.3, 37.4, 45.8, 68.8, 84.6 to 105 kPa were considered to assess the reproducibility and the stiffness sensitivity of these methods as illustrated in Fig. 3a.

**Depth discrimination using homogeneous phantoms.** For this second protocol, the objective was to estimate the sensitivity of the five techniques to changes of the phantom thickness. To do so, four polyurethane gel-based phantoms with thicknesses $h$ of 1, 3, 6 and 10 mm were considered for each of the six Young's modulus values $E$ fixed in the range from 30.3 to 105 kPa as illustrated in Fig. 3b.

**Axial sensitivity using bilayer phantoms.** For this third protocol, the aim was to estimate the sensitivity of the five techniques to changes the properties (thickness and Young's modulus) of the upper layer in a bilayer configuration. Two bilayer configurations were considered with a 10 mm thick polyurethane phantom for the bottom layer ($h_l$ = 10 mm) having a Young's modulus $E_l$ of 84.6 kPa, as illustrated in Fig. 3c. In the first configuration, the Young's modulus $E_{up}$ of the upper layer was equal to 45.8 kPa and the thickness $h_{up}$ varied between 1 and 10 mm. This configuration will be referred to as "soft on rigid" in what follows. In the case of the second configuration, the Young's modulus $E_{up}$ of the upper layer was equal to 216 kPa and the thickness $h_{up}$ also varied between 1 and 10 mm. This configuration will be referred to as "rigid on soft" in what follows.

**Data analysis.** For the three protocols described above, the values of the various indicators were summarized with their means and standard deviations for all configurations. Statistical analyses were then performed to identify significant variations of the different indicators. The normality of the data was checked using the Shapiro-Wilk test. In the case of a normal data distribution, parametric Student's T-tests were performed to highlight a significant difference between the means of two groups of measurements. For other distributions, non-parametric Mann–Whitney U tests were also performed, with the same objective of identifying significant differences. For all comparisons, the significance level was fixed at 5%, the one-tailed option was applied, with the left or right direction chosen on the basis of general tendencies of indicators.

# 3. Results

## 3.1 Stiffness discrimination using homogeneous phantoms

Figure 4 shows the variations of the indicators $\Delta t$, $Ind$, $R_0$, $S$ and $H$ obtained respectively with IBAM, IndentoPro, Cutometer, MyotonPro and Durometer as a function of the Young's modulus $E$ for the homogeneous phantoms with a thickness of 10 mm. $\Delta t$ and $R_0$ (respectively $Ind$, $S$ and $H$) are shown to decrease (respectively increase) as a function of the Young's modulus. IBAM and IndentoPro allow to significantly discriminate the six phantoms according to their stiffness. In contrast, Cutometer, Durometer and MyotonPro can significantly differentiate only certain phantoms according to their stiffness.

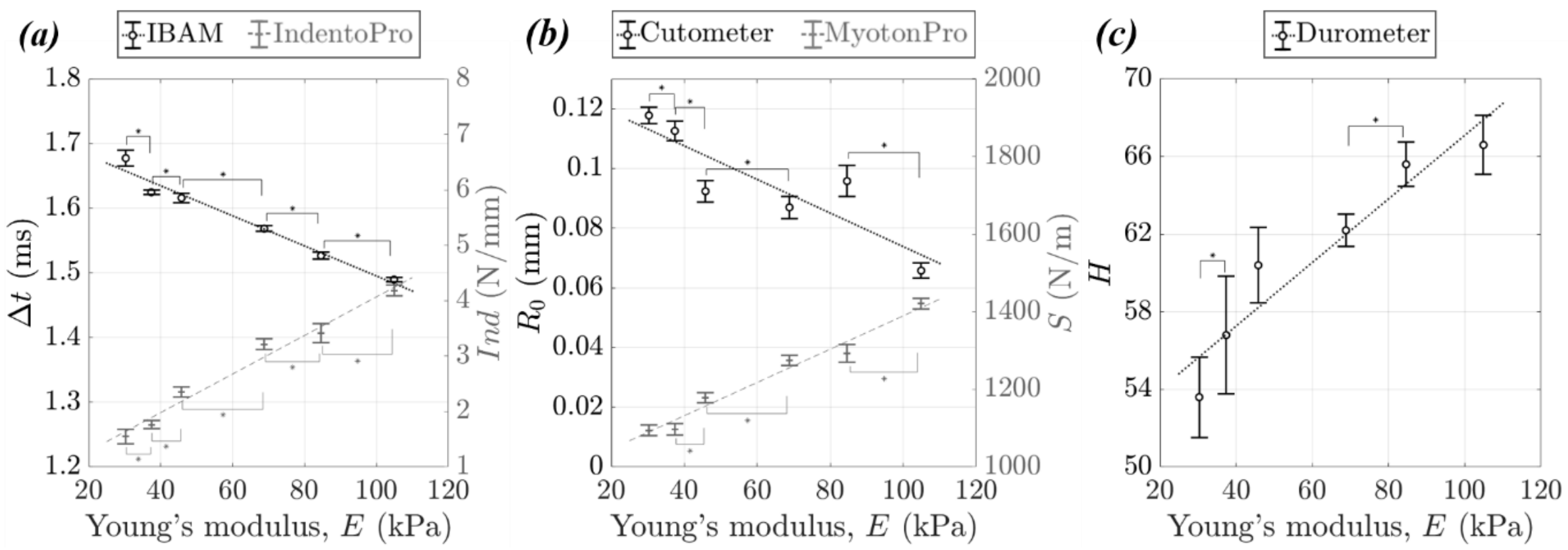

***Figure 4:*** *Variation of the indicators for the different techniques as a function of Young's modulus* $\boldsymbol{E}$ *of homogeneous phantoms:* ***(a)*** $\boldsymbol{\Delta t}$ *and* $\boldsymbol{Ind}$ *obtained with the IBAM and IndentoPro,* ***(b)*** $\boldsymbol{R_0}$ *and* $\boldsymbol{S}$ *obtained with the Cutometer and MyotonPro and* ***(c)*** $\boldsymbol{H}$ *obtained with the Durometer. The markers (respectively the error bars) correspond to the mean (respectively the standard deviation) of the measurements. * denotes significant difference (p-value < 0.05) between the values.*

The dashed and dotted lines correspond to linear regressions of the indicators of the different techniques as a function of $E$. For all techniques, the values of the regression analysis follow the model:

$$y = a \times E + b \qquad (2)$$

with $y$ corresponding to the different indicators. The values of $a$, $b$ and of the determination coefficient $R^2$ are summarized in Table 2 and will be exploited in the subsection "Stiffness discrimination using homogeneous phantoms" of the section 4.2 in the Discussion.

| **Mechanical surface characterization techniques** | $\boldsymbol{a}$ | $\boldsymbol{b}$ | $\boldsymbol{R^2}$ |
|---|---|---|---|
| IBAM | -0.0023 *ms.kPa*$^{-1}$ | 1.7 *ms* | 0.97 |
| IndentoPro | 0.035 *N.mm*$^{-1}$*.kPa*$^{-1}$ | 0.59 *N.mm*$^{-1}$ | 0.98 |
| Cutometer | -0.00056 *mm.kPa*$^{-1}$ | 0.13 *mm* | 0.77 |
| MyotonPro | 4.3 *N.m*$^{-1}$*.kPa*$^{-1}$ | 960.5 *N.m*$^{-1}$ | 0.97 |
| Durometer | 0.16 *kPa*$^{-1}$ | 51 | 0.90 |

***Table 2:*** *Results of the linear regression analysis of each indicator* $\boldsymbol{y}$ *as a function of the Young's modulus* $\boldsymbol{E}$ *for each technique (* $\boldsymbol{a}$ *the slope,* $\boldsymbol{b}$ *the intercept and* $\boldsymbol{R^2}$ *the coefficient de determination).*

### 3.2 Measuring depth using homogeneous phantoms

Figure 5 shows the variation of the indicators $\Delta t$, $Ind$, $R_0$, $S$ and $H$ obtained respectively with IBAM, IndentoPro, Cutometer, MyotonPro and Durometer as a function of the thickness $h$ of two homogeneous phantoms with Young's modulus of 37.4 or 84.6 kPa. $\Delta t$ (respectively $Ind$, $S$ and $H$) is shown to significantly increase (respectively decrease) as a function of the thickness $h$. $R_0$ does not show any clear behavior as a function of *h* for the two values of *E*.

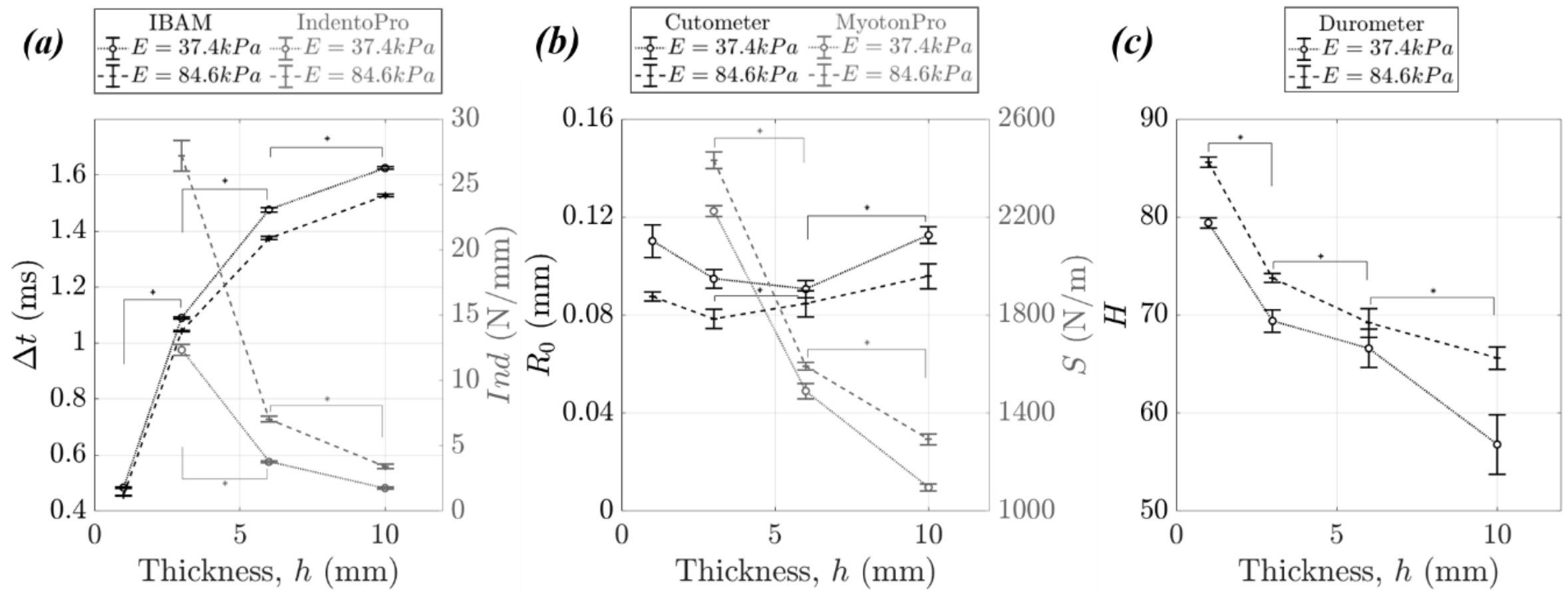

***Figure 5:*** *Variation of the indicators for the different techniques as a function of the thickness* $\boldsymbol{h}$ *of two homogeneous phantoms (*$\boldsymbol{E}$*=37.4 or 84.6 kPa):* ***(a)*** $\boldsymbol{\Delta t}$ *and* $\boldsymbol{Ind}$ *obtained with the IBAM and IndentoPro,* ***(b)*** $\boldsymbol{R_0}$ *and* $\boldsymbol{S}$ *obtained with the Cutometer and MyotonPro and* ***(c)*** $\boldsymbol{H}$ *obtained with the Durometer. The markers (respectively the error bars) correspond to the mean (respectively the standard deviation) of the measurements. * denotes significant difference (p-value < 0.05) between the values.*

### 3.3 Axial sensitivity using bilayer phantoms

Figure 6 shows the variation of the indicators $\Delta t$, $Ind$, $R_0$, $S$ and $H$ obtained respectively with IBAM, IndentoPro, Cutometer, MyotonPro and Durometer as a function of the upper layer thickness $h_{up}$ of two configurations of bilayer phantom (the “soft on rigid” configuration with $E_{up}$ = 45.8 kPa and the “rigid on soft” configuration with $E_{up}$ = 216.2 kPa). In the case of the “soft on rigid” configuration, $\Delta t$ (respectively $Ind$, $S$ and $H$) is shown to significantly increase (respectively decrease) as a function of $h_{up}$. In the other case of the “rigid on soft” configuration, $\Delta t$ (respectively $Ind$ and $H$) is shown to significantly decrease (respectively increase) as a function of $h_{up}$, except for $h_{up}$ > 3 mm, where $H$ stays constant. In contrast, for the two configurations, $R_0$ doesn't show any clear behavior as a function of $h_{up}$.

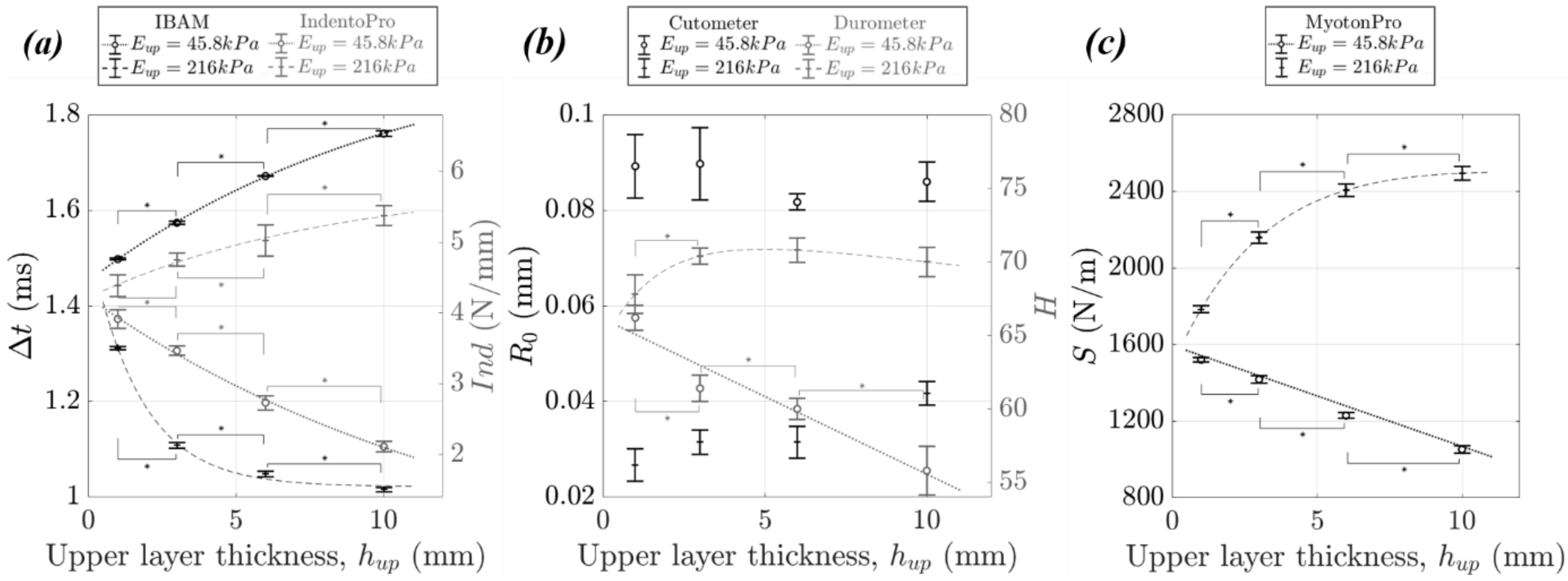

***Figure 6:*** *Variation of the indicators for the different techniques as a function of upper layer thickness* $\boldsymbol{h_{up}}$ *of two bilayer phantoms: the “soft on rigid” configuration with* $\boldsymbol{E_{up}}$*=45.8 kPa and the “rigid on soft” configuration with* $\boldsymbol{E_{up}}$*=216 kPa:* ***(a)*** $\boldsymbol{\Delta t}$ *and* $\boldsymbol{Ind}$ *obtained with the IBAM and IndentoPro,* ***(b)*** $\boldsymbol{R_0}$ *and* $\boldsymbol{H}$ *obtained with the Cutometer and Durometer and* ***(c)*** $\boldsymbol{S}$ *obtained with the MyotonPro. The markers (respectively the error bars) correspond to the mean (respectively the standard deviation) of the measurements. * denotes significant difference (p-value < 0.05) between the values.*

The dashed and dotted lines correspond to regression analysis of the indicators of the different techniques as a function of $h_{up}$ only for situations that demonstrate a significant behavior, leading to the estimated regression equations between $y$ and $h_{up}$ (with $y$ corresponding to the indicators of IBAM, IndentoPro, MyotonPro and Durometer). The values of regression analysis are summarized in Table 3 and will be exploited in the subsection “Axial sensitivity using bilayer phantoms” of the section 4.2 in the Discussion.

| **Mechanical surface characterization techniques** | | **Estimated regression equations** | $R^2$ |
|---|---|---|---|
| IBAM | *"soft on rigid"* | $\Delta t = -0.51 \times e^{-0.092\, h_{up}} + 2.0$ | 0.99 |
| | *"rigid on soft"* | $\Delta t = 0.52 \times e^{-0.58\, h_{up}} + 1.0$ | 0.99 |
| IndentoPro | *"soft on rigid"* | $Ind = 4.1 \times e^{-0.073\, h_{up}} + 0.11$ | 0.99 |
| | *"rigid on soft"* | $Ind = -1.6 \times e^{-0.12\, h_{up}} + 5.9$ | 0.99 |
| Cutometer | / | / | / |
| MyotonPro | *"soft on rigid"* | $S = 1592 \times e^{-0.042\, h_{up}}$ | 0.99 |
| | *"rigid on soft"* | $S = 2645 \times e^{-0.004\, h_{up}} - 1161 \times e^{-0.31\, h_{up}}$ | 1 |
| Durometer | *"soft on rigid"* | $H = 15.4 \times e^{-0.14\, h_{up}} + 52.4$ | 0.96 |
| | *"rigid on soft"* | $H = 73 \times e^{-0.004\, h_{up}} - 8.1 \times e^{-0.58\, h_{up}}$ | 1 |

***Table 3:*** *Results of the regression analysis of each indicator as a function of the upper layer thickness* $\boldsymbol{h_{up}}$ *for each technique (with the estimated equations and* $\boldsymbol{R^2}$ *the coefficient de determination).*

# 4. Discussion

The originality of this paper was to compare several multiphysical mechanical surface characterization techniques based on the indentation with IndentoPro and Durometer, suction with Cutometer, damped vibration with MyotonPro and impact analysis with IBAM. The choice of a synthetic material allowed to minimize sources of uncertainty on biological tissues (*ex vivo* or *in vivo*) and ensure relative stability of mechanical properties within the soft tissue mimicking phantom (compared to agar-based samples [41]).

## 4.1 Choice of parameters

**General.** In this comparative study, each mechanical surface characterization technique was repeated five times for all soft tissues mimicking phantom configurations. The choice of 5 repetitions was made as a compromise between a sufficient number to assess the reproducibility reliably and a reasonable duration of the experiments.

**IBAM.** As demonstrated in Poudrel et al. [24], the value of the indicator $\Delta t$ decreases as a function of the impact force $IF$. Thus, following the results obtained in Bouffandeau et al. [25], the range of acceptable $IF$ (5N : [30;35N]) was chosen to find a compromise between a sufficiently large range to obtain acceptable impacts in a reasonable time (here, 2/3 of all impacts were acceptable) and a sufficiently small range to avoid excessive effect of the IF on the results.

**Cutometer.** The interpretation of the curve representing the variation of the deformation as a function of time obtained by a suction with Cutometer provides multiple parameters describing the viscoelastic behavior of the soft tissues. Only the parameter $R_0$, defined as the final distension of the tissue at the end of the aspiration cycle, is related to the resistance of the tissue to the suction force [42].

**MyotonPro.** For each measurement, MyotonPro provides 5 output parameters related to the biomechanical properties of the soft tissue samples (the oscillation frequency, the dynamic stiffness, the logarithmic decrement, the mechanical stress relaxation time and the ratio of relaxation and deformation time). As our mechanical parameter of interest is the Young's modulus, our choice was to focus on the parameter $S$, defined by the manufacturer as the resistance of the tissues to a force of deformation produced by the mechanical impulse [40]. In a previous paper [25], the relationship between the dynamic stiffness $S$ and the Young's modulus of soft tissues has already been validated using agar-based soft tissues phantoms.

### 4.2 Calculation method for estimation errors

The objective of this study was to compare the performance of several mechanical surface characterization techniques. To do so, the uncertainties on the estimation of $E$ and $h$ in the homogeneous cases and of $h_{up}$ for both configurations "soft on rigid" and "rigid on soft" in the bilayer case, were calculated using the following approach for each measurement technique. The determination of the error $\sigma_x$ on the estimation of $x$ (with $x$ corresponding to $E$, $h$ or $h_{up}$ according to the different protocols of the study) was defined by [43]:

$$\sigma_x = \left| \left(\frac{dy}{dx}\right)^{-1} \times \sigma_y \right|, \quad (3)$$

where y is the considered indicator, $\sigma_y$ is its estimation error given by its standard deviation, and $\frac{dy}{dx}$ is the derivative of $y$ with respect to $x$, which is estimated using the regression analyses reported in Tables 2 and 3.

**Stiffness discrimination using homogeneous phantoms.** Here, $x = E$, $\frac{dy}{dE} = a$ (see Table 2) and the standard deviation $\sigma_y$ corresponds to the average values of the standard deviation of $y$ for $E \in [30.3; 105]$ kPa. The results are given in Table 4.

| **Mechanical surface characterization techniques** | $\frac{dy}{dE}$ | $\sigma_y$ | $\sigma_E$ |
|---|---|---|---|
| IBAM: $y = \Delta t$ | -2.3 $\mu s.kPa^{-1}$ | 6.1 $\mu s$ | 2.6 $kPa$ |
| IndentoPro: $y = Ind$ | 0.035 $N.mm^{-1}.kPa^{-1}$ | 0.11 $N.mm^{-1}$ | 3.1 $kPa$ |
| Cutometer: $y = R_0$ | -0.56 $\mu m.kPa^{-1}$ | 3.5 $\mu m$ | 6.2 $kPa$ |
| MyotonPro: $y = S$ | 4.3 $N.m^{-1}.kPa^{-1}$ | 15 $N.m^{-1}$ | 3.5 $kPa$ |
| Shore Durometer: $y = H$ | 0.16 $kPa^{-1}$ | 1.8 | 11 $kPa$ |

***Table 4:*** *Results of the uncertainties on the estimation of the Young's modulus* $\boldsymbol{\sigma_E}$ *with homogeneous soft tissue mimicking phantoms for each indicator.* $\frac{dy}{dE}$ *is the derivative of* $\boldsymbol{y}$ *with respect to* $\boldsymbol{E}$ *and* $\boldsymbol{\sigma_y}$ *is the standard deviation of* $\boldsymbol{y}$.

As shown in Table 4, IBAM, MyotonPro and IndentoPro demonstrate the best sensitivity regarding the stiffness of a homogeneous soft tissue mimicking phantom, which is indicated by the lowest values for $\sigma_E$, the uncertainty on the estimation of the Young's modulus $E$. This performance of IBAM can be explained by the low uncertainties of the measurements, with standard deviations of around 0.4% relatively to the mean values of $\Delta t$. The other methods have higher reproducibilities from 1.2% for MyotonPro to 4.4% for IndentoPro. Note that $\Delta t$, $S$ and $Ind$ vary almost linearly as a function of $E$ ($R^2$ = 0.97-0.98, Table 2) while for the other methods (an in particular for Cutometer), the indicators vary nonlinearly, which reinforces the better performances of IBAM, MyotonPro and IndentoPro compared to the 2 others techniques.

**Measuring depth using homogeneous phantoms.** Here, $x = h$ and $\sigma_h$ corresponds to the average values of the standard deviation of $h$ for each $E \in$ [30.3; 105] kPa. For each value of $E$, the derivatives $\frac{dy}{dh}$, approximated with linear regressions of $y$ as a function of $h$, and standard deviations $\sigma_y$ were estimated for each $h \in$ [1; 10] mm. The results are given in Table 5 for $E$ = 37.4 and $E$ = 84.6 kPa.

| Mechanical surface characterization techniques | | $\frac{dy}{dh}$ | $\sigma_y$ | $\sigma_h$ |
|---|---|---|---|---|
| IBAM - $\Delta t$ | *37.4 kPa* | 119 *μs.mm*$^{-1}$ | 4.3 *μs* | 0.036 *mm* |
| | *84.6 kPa* | 111 *μs.mm*$^{-1}$ | 3.8 *μs* | 0.034 *mm* |
| IndentoPro - $Ind$ | *37.4 kPa* | -1.46 *N.mm*$^{-2}$ | 0.19 *N.mm*$^{-1}$ | 0.13 *mm* |
| | *84.6 kPa* | -3.26 *N.mm*$^{-2}$ | 0.52 *N.mm*$^{-1}$ | 0.16 *mm* |
| Cutometer - $R_0$ | *37.4 kPa* | 0.0005 | 4.3 *μm* | 8.5 *mm* |
| | *84.6 kPa* | 0.0012 | 4.1 *μm* | 3.3 *mm* |
| MyotonPro - $S$ | *37.4 kPa* | -158 *kN.m*$^{-2}$ | 22 *N.m*$^{-1}$ | 0.14 *mm* |
| | *84.6 kPa* | -158 *kN.m*$^{-2}$ | 24 *N.m*$^{-1}$ | 0.15 *mm* |
| Durometer - $H$ | *37.4 kPa* | -2.3 *mm*$^{-1}$ | 1.7 | 0.73 *mm* |
| | *84.6 kPa* | -2.0 *mm*$^{-1}$ | 0.9 | 0.45 *mm* |

***Table 5:*** *Results of the uncertainties on the estimation of the thickness* $\boldsymbol{\sigma_h}$ *with homogeneous soft tissue mimicking phantoms (*$\boldsymbol{E}$ *= 37.4 or 84.6 kPa) for each indicator.*

As shown in Table 5, IBAM has the best sensitivity regarding the depth estimation, which correspond to the lowest values of $\sigma_h$. While uncertainties of IndentoPro and MyotonPro are three to four times higher to that of the IBAM, these two devices still outperform Durometer and Cutometer for thicknesses between 1 and 10 mm. However, the characterization of phantoms with $h$ = 1 mm were not possible with IndentoPro and MyotonPro. In the case of Cutometer, its low sensitivity for $h$ > 1 mm can be explained by the physical phenomenon implemented in the device, suction, and the restricted aperture of the probe (2 mm diameter in our study). According to the manufacturer [40] and our previous data [25], MyotonPro is sensitive to mechanical properties down to a depth of around 20 mm, so this technique is mainly used to investigate subcutaneous regions such as superficial muscles and tendons.

**Axial sensitivity using bilayer phantoms.** Here, $x = h_{up}$ and nonlinear regressions were performed to study the variation of *y* as a function of *x* . Thus, it was decided to estimate the $\frac{dy}{dx}$ and $\sigma_x$ for both configurations for $x$ = 5 mm, using the equations shown in Table 3. In addition, the standard deviations $\sigma_y$ were averaged for $x \in [3 ; 10]$ mm. The results are given in Table 6.

| Mechanical surface characterization techniques | | $\frac{dy}{dh_{up}}$ | $\sigma_y$ | $\sigma_{h_{up}}$ |
|---|---|---|---|---|
| IBAM - $\Delta t$ | *“soft on rigid”* | 30 $\mu s.mm^{-1}$ | 3.5 $\mu s$ | 0.12 *mm* |
| | *“rigid on soft”* | -17 $\mu s.mm^{-1}$ | 5.2 $\mu s$ | 0.31 *mm* |
| IndentoPro - $Ind$ | *“soft on rigid”* | -0.21 $N.mm^{-2}$ | 0.080 $N.mm^{-1}$ | 0.38 *mm* |
| | *“rigid on soft”* | 0.11 $N.mm^{-2}$ | 0.15 $N.mm^{-1}$ | 1.44 *mm* |
| Cutometer - $R_0$ | / | / | / | / |
| MyotonPro - $S$ | *“soft on rigid”* | -54 $kN.m^{-2}$ | 17 $N.m^{-1}$ | 0.32 *mm* |
| | *“rigid on soft”* | 66 $kN.m^{-2}$ | 32 $N.m^{-1}$ | 0.49 *mm* |
| Durometer - $H$ | *“soft on rigid”* | -1.1 $mm^{-1}$ | 1.1 | 1.00 *mm* |
| | *“rigid on soft”* | 0.39 $m^{-1}$ | 0.79 | 2.1 *m* |

***Table 6:*** *Results of the uncertainties on the estimation of the upper layer thickness* $\boldsymbol{\sigma_{h_{up}}}$ *with bilayer soft tissue mimicking phantoms for each indicator.* $\frac{dy}{dh_{up}}$ *is the derivative of* $\boldsymbol{y}$ *with respect to* $\boldsymbol{h_{up}}$ *and* $\boldsymbol{\sigma_y}$ *is the standard deviation, where* $\boldsymbol{y}$ *correspond to* $\boldsymbol{\Delta t}$, $\boldsymbol{Ind}$, $\boldsymbol{R_0}$, $\boldsymbol{S}$ *or* $\boldsymbol{H}$.

As shown in Table 6, for the “soft on rigid” and “rigid on soft” configurations, the sensitivity to the estimation of $h_{up}$obtained with IBAM corresponds to the lowest values of $\sigma_{h_{up}}$. Among the other methods, MyotonPro and IndentoPro are the only ones reaching acceptable values of $\sigma_{h_{up}}$, despite uncertainties 2.5 to 4.5 times higher than with IBAM. As in the subsection about the stiffness discrimination, this performance of IBAM can be explained by the low uncertainties of the measurements, with standard deviations of around 0.3% relative to the mean values of the indicator for IBAM, compared with other methods, which have higher reproducibility from 1.3% to 3.1%.

For the "rigid on soft" configuration, a degradation of the sensitivity to the estimation of $h_{up}$ can be noted for all techniques compared with the "soft on rigid" configuration. Only IBAM and MyotonPro reached acceptable results. As previously mentioned, Cutometer is not sensitive to mechanical properties for $h > 1$ mm in both bilayer phantom configurations.

### 4.3 Comparison with the literature

Our results are qualitatively consistent with previous studies. According to our study with agar-based soft tissues phantoms [25], for $E \in [20 ; 100]$ kPa, the sensitivity to stiffness variation of the previous experimental version of IBAM was equivalent to that of MyotonPro. However, the axial sensitivity of MyotonPro using bilayer phantoms was two times lower than that of IBAM. In the present study, the stiffness sensitivity of the current version of IBAM including the improvements was improved and it is now 1.3 times better compared to MyotonPro.

The results found herein corresponding to the comparison between several techniques to assess soft tissue properties may be compared with results obtained from the literature. Cutometer is a device widely used to study the elastic properties of the skin [44] in dermatology, such as with scleroderma [45] or wound healing [3], and also in cosmetics to assess the aging of the skin [9]. Various studies have demonstrated the consistency of the Cutometer with other methods such as the SkinFibrometer® (Delfin Technologies, Kuopio, Finland) [46], an indentation technique, and also with the Ballistometer® (Diastron Ltd., Andover, UK) [44], a damped vibration analysis technique, for characterizing the biomechanical behavior of skin. Although SkinFibrometer® and Ballistometer® appeared to be more sensitive to stiffness than to elastic properties of the tissue [44,46]. However, the probe with a 2 mm diameter aperture was developed to perform the characterization of the epidermis of the skin with Cutometer [3]. Therefore, variations located more than 10 mm below the surface of the phantom were not detected in this comparative study. A change of probe for the 6 mm diameter aperture, designed to characterize the dermis [3], could allow an estimation of axial sensitivity with bilayer phantoms. According Dellalana et al. [7], the inter-observer reproducibility of MyotonPro was 1.2 times better than that of Durometer and exhibited sensitivity to stiffness variation with a normalized minimal detectable change around 7.6% for *in vivo* healthy human skin measurements. Bartsch et al. [22] showed that IndentoPro, MyotonPro and Durometer were sensitive to stiffness variations in the layers (with thickness greater than 1 mm) of a multi-layered soft tissue phantom composed of the same material as ours. In addition, IndentoPro and MyotonPro could detect mechanical property variations located more than 10 mm below the surface of the phantom. In this same study, strain ultrasound elastography revealed limitations in the characterization of mechanical properties located less than 3 mm below the surface. Pouletaut et al. [47] confirmed this observation by pointing out the absence of wave propagation in the material due to its homogeneity. However, other forms or types of elastographic technique may seem better adapted to the skin, such as shear wave elastography [16] or OCE [48]. Elastographic techniques are among the few devices available for assessing the mechanical properties of biological tissues. Thus, clinical approaches have been and/or are being developed in dermatology and aesthetic medicine. Depending on the pathology, skin lesions may appear much stiffer than healthy or benign tissue (e.g. malignant neck tumors were 8 times stiffer than benign tumors [49]) and less elastic (e.g. higher deformations in malignant melanomas than healthy tissue [50]). Nevertheless, despite more complex data acquisition enabling spatial stiffness cartography, elastography techniques do not represent the same investment in terms of cost or ease of use as the five mechanical surface characterization techniques in this comparative study.

Note that for MyotonPro, an alternative probe, called “L-shape”, is proposed in order to reduce the measurement depth and estimate mechanical properties more superficially, such as for skin [51]. With the standard probe, MyotonPro is positioned perpendicular to the tissue, whereas with the “L-shape” probe, this same device is positioned parallel and generates a mechanical impulse that is also parallel to the surface of the tissue. Thus, with this other probe, the nature of the mechanical excitation would be reconsidered as shear [52], and therefore comparisons with other techniques would be more difficult, as the measurements would be correlated to different properties (Young's modulus vs. Shear modulus).

### 4.4 Limitations and perspectives

Polyurethane gel pads (Technogel Germany GmbH, Berlingrode, Germany) were used for the soft tissue mimicking phantoms. This pre-prepared pads were chosen for their excellent long-term conservation, homogeneity of properties and simplicity of use (no preparation is required in advance) compared to agar-based materials [37,41]. However, this type of material also has certain disadvantages. First, the Young's modulus values of the different gel pads are obtained from manufacturing data, since no mechanical tests (traction, compression, etc.) could be performed to avoid damaging to the phantoms. Moreover, the phantoms were thin plates, which makes the measurement of their actual mechanical properties using classical mechanical testing devices difficult. Therefore, the values of the phantom Young's modulus, and subsequently the estimation errors, may be biased because these data, which are provided by the manufacturer, could not been checked. However, this bias remained constant when different mechanical surface characterization techniques were used to conduct measurements with the same phantom. Additionally, the absence of independent validation prevents us from quantitatively comparing our data (e.g., stiffness or axial sensitivity) with literature-documented values. However, the stiffness range of these phantoms is consistent with the data from the literature, which estimates Young's modulus of soft biological tissues between 1 and 200 kPa [53,54] and others soft tissues mimicking phantoms made with agarose [25] or plastisol [55,56]. Second, we were limited by the available samples for the range of variation of the relevant parameters ($E$, $h$ and $h_{up}$). Therefore, it was not possible to refine our experimental configurations using intermediate or additional phantoms. In particular, with the MyotonPro and IndentoPro, it would have been interesting to investigate the results for h > 10 mm, to compensate for the difficulties encountered with $h$ = 1 mm. Third, unlike biological tissues, which have bonded layers, the bilayer phantoms of this study exhibit an imperfect interface between two media, with a very low friction coefficient, leading to a weak transmission of shear stresses and this interface. Multilayer configurations of these phantoms had already been used to mimic soft tissue structures in the human lumbar region [22]. Such phantoms have proved to be adapted for compressive stiffness measurements [22]. Other forms of bilayer phantoms were also investigated (data not shown), by considering a thin layer of double-sided adhesive tape located between the two phantom layers, leading to different results. The presence of an adhesive tape between the two layers of the bilayer phantom alters the stress distribution during measurements due to the change in boundary conditions, which emphasizes the importance of interface bounding. Moreover, a polyurethane film covers the gel and, despite its very low thickness of 25 µm, it may influence the measurements. Fourth, the material of the phantoms was simpler than biological soft tissues, which are heterogeneous and anisotropic. Additionally, this comparative study focused solely on the stiffness of the phantoms. However, biological soft tissues appear to exhibit more complex behavior, including viscous and hyperelastic effects. This simplicity allows us to work under standardized conditions and to achieve a reliable comparison of the different measurement techniques. Therefore, it would be interesting to conduct parallel measurements with IBAM and other techniques in future *ex vivo* and *in vivo* studies to validate the findings of this paper.

The IBAM project is currently in its research phase. For this reasons, various technical solutions are being investigated to miniaturize the experimental set-up into an easy-to-use, handheld and reliable prototype, which could pave the way for developing a future medical device. First, the automation of the measurement must be considered, which is outside the scope of the present study. Hammer impacts could for instance be generated using an actuator. The force sensor, currently located in the hammer head, could for example be attached at the end of the actuator shaft. This implementation should enable full automation of the measurement process, thus minimizing human interference. However, attention to adjusting different sequences of the measurement process (e.g. impact, recording, rest) will be required. Second, certain aspects of the prototype may require further investigation regarding in vivo measurements on biological tissue. For instance, the measurement depth may be adjusted by redesigning the punch geometry or adjusting the impact force interval. Third, artificial intelligence tools could also improve the interpretation of force signals. Numerical model should also be developed to provide more physical insight on the phenomena occurring during impacts. Eventually, in light of these previous *in vitro* studies with soft tissue mimicking phantoms, future studies should be conducted with biological tissues from animal models (both *ex vivo* and *in vivo*) in order to validate the IBAM approach in a configuration closer to the clinical situation. Ex vivo and in vivo studies are mandatory to demonstrate the clinical applicability of IBAM prior to the initiation of clinical studies in human subjects.

## 5. Conclusion

In this comparative study of mechanical surface characterization techniques, IBAM was identified as the best choice for the characterization of mechanical properties of soft tissues phantoms at relatively low depth (between 1 and 10 mm), followed by MyotonPro and IndentoPro. In terms of stiffness and thickness sensitivities with homogeneous and bilayer phantoms, the performance of IBAM is between 1.2 and 4.5 times better than that of MyotonPro and IndentoPro. In comparison, the performance of Durometer is around 10 times lower than that of IBAM. Finally, Cutometer appears to be the least relevant technique for measuring and detecting any changes in superficial tissue condition between 1 and 10 mm depth. However, the clinical applicabilities of IBAM remains to be demonstrated and the findings of this study must be confirmed by conducting *ex vivo* and *in vivo* studies on biological tissues.

## 6. Funding

This project has received funding from the European Research Council (ERC) under Horizon 2020 (grant agreement #101147066, project ERC Proof of Concept AssesSkin), from the project OrthAncil (ANR-21-CE19-0035-03), from the project OrthoMat (ANR-21-CE17-0004) and from the project MoDyBe (ANR-23-CE45-0011). The authors acknowledge the support from CNRS Innovation for the Prematuration project CarImpact. The authors acknowledge the support of Suzanne De Vuyst for performing some experimental measurements.

## 7. Authors' Contributions

## 8. Declaration of Conflicting Interests

The authors declare that they have no financial or non-financial interests that are directly or indirectly related to the work submitted for publication.

## 9. Ethical Statement

This study did not require any ethical approval, as no human nor animal subjects were involved.